\documentclass[12pt]{article}
\usepackage{a4}
\usepackage{graphics}
\usepackage{graphicx}

\input epsf.sty

\def\slashchar#1{\setbox0=\hbox{$#1$}
   \dimen0=\wd0 \setbox1=\hbox{/} \dimen1=\wd1
   \ifdim\dimen0>\dimen1 \rlap{\hbox to \dimen0{\hfil/\hfil}} #1
   \else  \rlap{\hbox to \dimen1{\hfil$#1$\hfil}} / \fi}

\begin{document}

\title{Application of chiral quark models to high-energy processes% 
\footnote{Presented at Bled 2004: Quark Dynamics, Bled (Slovenia), 12-19 July  2004}}

\date{September 2004}

\author{\underline{Wojciech Broniowski$^1$} and Enrique Ruiz Arriola$^2$\\
{\small $^1$ The H. Niewodnicza\'nski Institute of Nuclear Physics}\\ 
{\small Polish Academy of Sciences, PL-31342 Cracow, Poland}\\
{\small $^2$Departamento de F\'{\i}sica Moderna, Universidad de Granada} \\ 
{\small E-18071 Granada, Spain}}

\maketitle

\abstract{We discuss the predictions of chiral quark models for basic
pion properties entering high-energy processes: generalized parton
distributions (GPD's) and unintegrated parton distributions
(UPD's). We stress the role of the QCD evolution, necessary to compare
the predictions to data.}

\bigskip

This is a very brief account of the talk based on
Refs.~\cite{impact,solkw,PDA,SQM}, where the reader is referred to for
the details and references.  We discuss the use of low-energy chiral
quark models to compute low-energy matrix elements of hadronic
operators appearing in high-energy processes, in particular we
evaluate the {\em generalized} and {\em unintegrated} parton
distributions (GPD's and UPD's) of the pion in the Nambu--Jona-Lasinio
model and the Spectral Quark Model \cite{SQM}. We carry on the QCD
evolution, necessary when comparing the model predictions to data
obtained at much higher scales. 

The twist-2 GPD of the pion is defined as
\begin{eqnarray}
H(x,\xi, - {\bf{\Delta}}_ \perp^2 ) = 
%\int d^2 b 
\int \frac{dz^-}{4\pi} e^{i x p^+ z^- 
%+ {\bf{\Delta}}_\perp \cdot {\bf{b}} )
} \langle \pi^+ (p') | \bar q (0,
-\frac{z^-}{2} , 0) {\gamma^+} q (0, \frac{z^-}{2} , 0)
| \pi^+ (p) \rangle ,\nonumber
\end{eqnarray}
where the quark operator $q(z^+,z^-,z_\perp)$ is on the light cone
$z^2=0$ and the link operators $P \exp(i g \int_0^z dx^\mu A_\mu)$ are
implicitly present to ensure the gauge invariance (as usual we work in
the light cone gauge $A_+=0$).  A similar definition holds for the
gluon distribution.  In chiral quark models the evaluation of $H$ at
the leading-$N_c$ (one-loop) level is straightforward.  For the NJL
model with the Pauli-Villars regularization we get {\begin{eqnarray}
&& H_{\rm NJL}(x,0,- {\bf \Delta}_\perp^2) = \nonumber \\ && \left [
1+\frac{N_c M^2 (1-x)|{\bf \Delta}_\perp|}{4 \pi^2 f_\pi^2 s_i} \sum_i
c_i \log \left ( \frac{s_i+(1-x)|{\bf \Delta}_\perp|}{s_i-(1-x)|{\bf
\Delta}_\perp|} \right )\right ] \theta(x) \theta(1-x) , \nonumber \\
&& s_i=\sqrt{(1-x)^2{\bf \Delta}_\perp^2+4M^2+4\Lambda_i^2} ,
\nonumber
\label{eq:unint_ff4}
\end{eqnarray}}
where {$M$} is 
the constituent quark mass, {$\Lambda_i$} are the PV regulators, 
and {$c_i$} are suitable
constants.  For the simplest twice-subtracted case, explored below, one has,
for any regulated function $f$, the operational definition
{\begin{eqnarray}
\sum_i c_i f( \Lambda_i^2 ) = f(0) - f(\Lambda^2 ) + \Lambda^2 df(\Lambda^2 )/d\Lambda^2. \nonumber
\label{eq:PV2} 
\end{eqnarray} }
We use $M=280$~MeV and $\Lambda=871$~MeV, which yields the pion decay 
constant $f_\pi=93$~MeV. In the SQM the result is
\begin{eqnarray} 
H_{\rm SQM}(x,0,- {\bf \Delta}_\perp^2) = \frac{m_\rho^2 ( m_\rho^2 - (1-x)^2
{\bf \Delta}_\perp^2)} {( m_\rho^2 + (1-x)^2 {\bf
\Delta}_\perp^2)^2} \theta(x)\theta(1-x), \label{res:vmd} \nonumber
\end{eqnarray}
where $m_\rho$ is the mass of the $\rho$ meson.
We check that the pion electromagnetic form factor is 
\begin{eqnarray}
F_{\rm SQM}(t)=\int_0^1 dx H_{\rm SQM}(x,0,t) = {\frac{m_\rho^2}{m_\rho^2+t}}, \nonumber
\end{eqnarray}  
which is 
the built-in vector-meson dominance principle. For both models {$F(0)=1$} and 
{$H_{\rm SQM}(x,0,0)=\theta(x)\theta(1-x)$}.

Our next goal is to compare the results to the data from transverse
lattices \cite{Dalley}.  We pass to the impact-parameter space via the
Fourier-Bessel transformation, as well as carry the LO DGLAP
perturbative QCD evolution from the low model scale $Q_0$=313~MeV
\cite{zak} up to the scale of the data.
\begin{figure}
\begin{center}
\includegraphics[width=11.3cm]{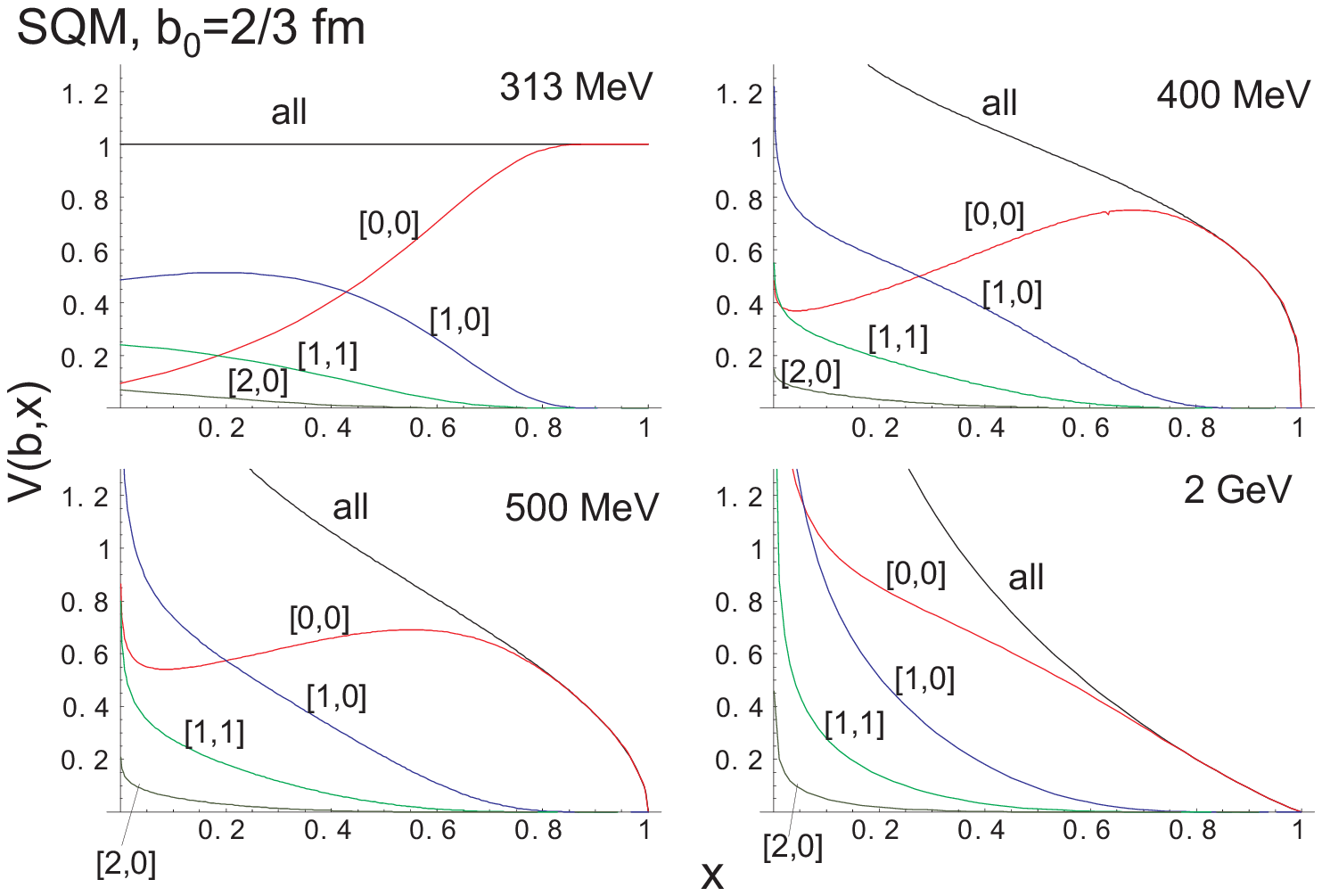}
\end{center} 
\vspace{-5mm}
\begin{center}
\includegraphics[width=8cm]{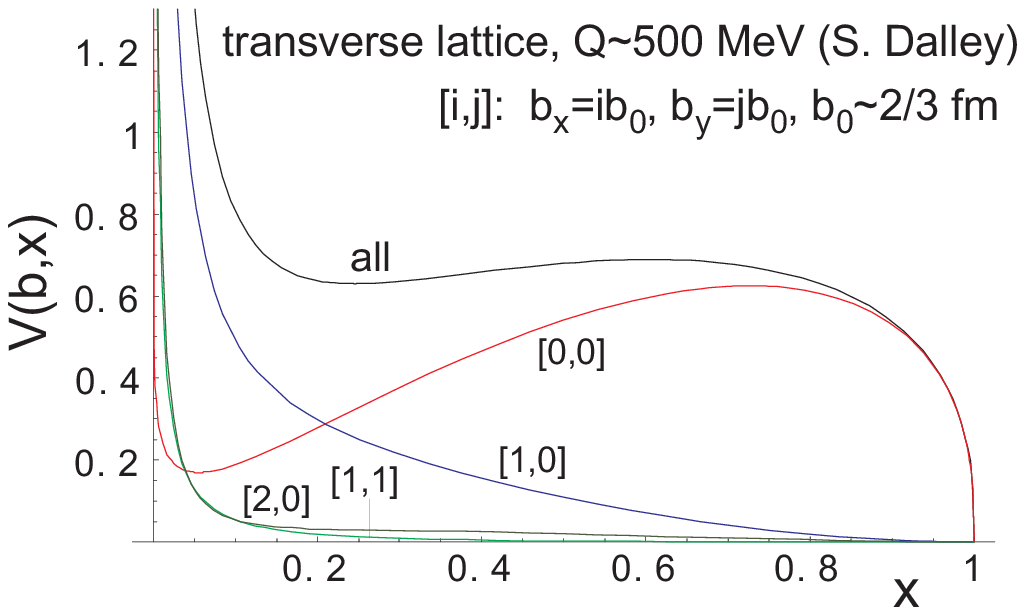}
\end{center}
\caption{GPD of the pion in the impact-parameter space plotted as a
function of the Bjorken $x$. Top: model for four momentum scales, from
313~MeV up to 2~GeV.  Bottom: transverse lattice
\cite{Dalley}. Numbers in brackets label the plaquette
\cite{impact}. The qualitative agreement to the data is achieved at
the scale of about 500~MeV. }
\label{bGPD}
\end{figure}
The results are shown in Fig.~\ref{bGPD}. We note that while the
results at $Q_0$ are completely different off the lattice data, when
evolved to the scale of 500~MeV, corresponding to the lattice
calculations, acquire a great resemblance to the data.

\begin{figure}
\begin{center}
\includegraphics[width=8cm]{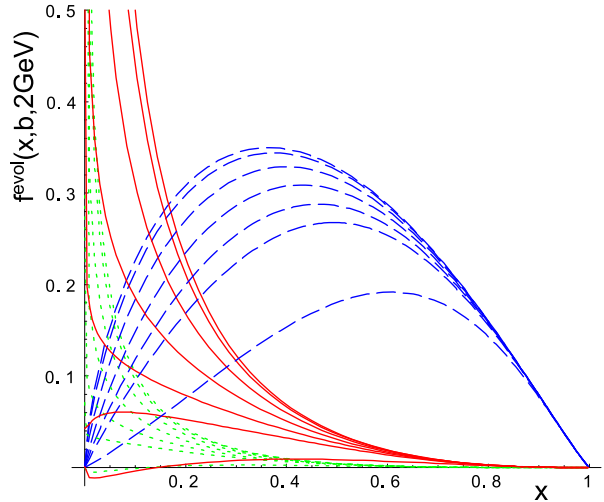}
\end{center}
\caption{Valence quarks (dashed lines), sea quarks (dotted lines), and
gluons (solid lines), for the transverse coordinate $b=0,1,2,3,4,5,$
and $10$~fm (bottom to top). Evolution with the Kwieci\'nski equations
from the model scale $Q_0$=313~MeV up to $Q=2$~GeV has been made.}
\label{q4} 
\end{figure}

In the second part of this talk we discuss the leading-twist 
UPD's of the pion, defined as
\begin{eqnarray}
q(x,{k_\perp})&=&\int \frac{dy^- d^2 y_\perp}{16\pi^3} e^{-ixp^+ y^-+i {k_\perp} \cdot y_\perp}
\langle p \mid \bar{\psi}(0,y^-,y_\perp) \gamma^+ \psi(0) \mid p \rangle , \nonumber
\end{eqnarray} 
and similarly for the gluon.
An elementary one-quark-loop calculation in the NJL model with the PV regularization gives
for $q$ and its Fourier-Bessel transform the result 
\begin{eqnarray}
q_{\rm NJL}(x,{k_\perp},Q_0)&=&\frac{\Lambda^4 M^2 N_c}
  {4 f_\pi^2 \pi^3 \left( {k_\perp^2} + M^2 \right) \left( {k_\perp^2} + \Lambda^2 + M^2 \right)^2} 
\theta(x)\theta(1-x) \nonumber \\
F^{\rm NP}_{\rm NJL}({b})&=&\frac{M^2 N_c}{4 f_\pi^2 \pi^2} \left( 2 K_0({b} M) - \label{qqNJLb2}
        2 K_0({b} \sqrt{\Lambda^2 + M^2}) - 
        \frac{{b} \Lambda^2 K_1({b} \sqrt{\Lambda^2 + M^2})}{\sqrt{\Lambda^2 + M^2}} \right ). \nonumber 
\end{eqnarray}
In SQM we find
\begin{eqnarray}
q_{\rm SQM}(x,{k_\perp},Q_0)&=&\frac{6m_\rho^3}{\pi({k_\perp^2}+m_\rho^2/4)^{5/2}}
 \theta(x)\theta(1-x), \nonumber \\
F^{\rm NP}_{\rm SQM}({b})&=&\left (1+\frac{{b} m_\rho}{2} \right ) \exp\left ( - \frac{m_\rho {b}}{2} 
\right ) \nonumber 
\end{eqnarray}
(the meaning of $b$ different here, it is the transverse coordinate
conjugated to $k_\perp$).  The above results are at the low model
scale $Q_0$.  Next, we evolve these UPD's from $Q_0$ to high scales
with the Kwieci\'nski equations \cite{solkw}, obtained in the CCFM
framework.  The results are displayed in Fig.~\ref{q4}.

One may show several qualitative and quantitative results concerning
UPD's.  At large $b$ they fall off exponentially and at large
$k_\perp$ they fall off as a power law.  Spreading with increasing
$Q^2$ occurs, with $\langle k_\perp^2 \rangle \sim Q^2 \alpha_S(Q^2)$.
Also, asymptotic formulas at limiting cases may be explicitly given
\cite{solkw} which may be useful in checking numerical calculations of
CCFM-type cascades \cite{Jung}.

Our basic conclusion is that chiral quark models may be used to
provide GPD's and UPD's (also the pion distribution amplitude
\cite{PDA} not presented here) at the low model scale, $Q_0$. Upon
evolution to higher scales, the agreement with the data (experimental
or lattice) is very reasonable.

\end{document}